\documentclass[11pt]{article}

\usepackage[margin=1.25in]{geometry}

\usepackage{amsmath,amssymb,amsfonts,amsthm,bbm}

\usepackage{graphicx}
\usepackage{booktabs}
\usepackage{array}
\usepackage{multirow}
\usepackage{makecell}
\usepackage{float}

\usepackage[authoryear]{natbib}
\usepackage[colorlinks=true,citecolor=blue,urlcolor=blue,linkcolor=blue]{hyperref}

\title{NFL step-and-turn: A generative framework for evaluating player movement in American football}

\author{
Quang Nguyen \qquad \qquad Ronald Yurko \bigskip \\
Department of Statistics \& Data Science, Carnegie Mellon University
}

\date{}

\begin{document}

\maketitle

\begin{abstract}
In sports analytics, player tracking data have driven significant advancements in the task of player evaluation. We present a novel generative framework  for evaluating the observed frame-by-frame player positioning against a distribution of hypothetical alternatives. We illustrate our approach by modeling the within-play movement of an individual ball carrier in the National Football League (NFL). Specifically, we develop Bayesian multilevel models for frame-level player movement based on two components: step length (distance between successive locations) and turn angle (change in direction between successive steps). Using the step-and-turn models, we perform posterior predictive simulation to generate hypothetical ball carrier steps at each frame during a play. This enables comparison of the observed player movement with a distribution of simulated alternatives using common valuation measures in American football. We apply our framework to tracking data from the first nine weeks of the 2022 NFL season and derive novel player performance metrics based on hypothetical evaluation.
\end{abstract}

\noindent%

\section{Introduction}
\label{sec:intro}

The emergence of player tracking data in recent years has fundamentally
reshaped the landscape of sports analytics. Many sports now collect
high-frequency spatiotemporal measurements of player locations
throughout a game. These data provide a level of granularity that
enables researchers to move from summary- and event-based analyses
toward more complex, continuous modeling of athletic performance and
game dynamics. For more detailed discussions on player tracking data in
sports, see \citet{albert2017handbook}, \citet{baumer2023big}, and
\citet{kovalchik2023player}.

In this work, we focus on a fundamental problem in sports analytics that
still remains open: player evaluation with tracking data. In particular,
we introduce a statistically-driven generative modeling framework for evaluating the
observed player positioning relative to a distribution of hypothetical movements at every moment within a play. We focus on
American football and leverage tracking data to assess the movement of
individual ball carriers on rushing plays. Below, we discuss the
football tracking data literature and highlight prior research on player
movement in sports that is relevant to our paper.

\subsection{Previous work: football tracking data}

In American football, the National Football League (NFL) collects player
tracking data via the Next Gen Stats system, which was launched in 2016.
The system employs radio frequency identification (RFID) chips installed
inside player shoulder pads and the football to record positional data
at a rate of 10 frames per second. This subsequently captures various
kinematic attributes (e.g., location, speed, acceleration, etc.) for
every player on the field during each play. Furthermore, to promote
public research, the NFL releases tracking data through its annual
analytics competition known as the Big Data Bowl
\citep{lopez2020bigger}. Since 2019, each edition of the competition has
featured a theme (e.g., special teams, linemen, tackling, pre-snap
motion, etc.) and provides participants with a sample of player tracking
data to explore football-specific questions related to that theme.

Over time, the NFL Big Data Bowl has facilitated a growing body of work
on statistical methods for player tracking data in American football.
For instance, using model-based clustering, \citet{chu2020route}
identify and characterize route types of receivers, whereas
\citet{dutta2020unsupervised} provide unsupervised labels for defensive
pass coverage schemes. In addition, \citet{burke2019deepqb} and
\citet{reyers2021quarterback} develop machine learning approaches for
assessing quarterback performance with tracking data. Moreover,
\citet{deshpande2020expected} model hypothetical catch probability on
passing plays, while \citet{yurko2020going} propose a framework for
within-play valuation of game outcomes.

Tracking data have also enabled in-game evaluation of specific abilities
in football such as pass rush \citep{nguyen2024here}, tackling
\citep{nguyen2025fractional}, and change of direction
\citep{nguyen2025bayesian}. Recently, \citet{nguyen2025multilevel} and
\citet{michels2026integrating} use pre-snap information to gain insights
into quarterback snap timing and pass coverage. It is important to
recognize that the aforementioned studies introduce assessment framework
for football aspects across the board, from offense to defense and even
pre-snap behavior. This marks a major step forward, as evaluations for
many of these areas were extremely limited before player tracking data
became publicly accessible.

\subsection{Previous work: movement models and hypothetical
evaluation}\label{sec:related-work}

After tracking data first became available for research, it quickly
became apparent that movement models are an essential component of
within-play evaluation frameworks in sports analytics. In basketball,
\citet{cervone2016multiresolution} propose a stochastic process model
for the evolution of a basketball possession to estimate expected
possession value. This framework relies on models at two separate levels
of resolution: a microtransition movement model for all players and a
macrotransition model for possession-level events like passes, shots,
and turnovers. These sub-models can then be combined to estimate the
instantaneous value at every moment within a possession. Later on,
\citet{wu2018modeling} model offensive player movement in basketball to
predict frame-level player position on the court. In soccer,
\citet{fernandez2021framework} develop a similar framework to
\citet{cervone2016multiresolution} to obtain the instantaneous expected
value of soccer possessions. Specifically, this approach decomposes the
total value into sub-models of three primary on-ball actions: ball
drives, passes, and shots.

With the availability of tracking data, an appealing approach for
evaluating players based on within-play space-time information is
commonly known as \emph{ghosting}. This technique aims to model player
behavior by comparing the positioning and trajectory of an observed
player to a baseline, average player, i.e., the ``ghost''
\citep{lowe2013lights}. Early ghosting approaches mainly focus on deep
imitation learning and output point predictions for the movement pattern
of baseline-level players. \citet{le2017data} model where a soccer
defender should be at any moment based on league-average behavior, and
identify spatial locations that minimize the offense's scoring chance.
\citet{le2017coordinated} propose a coordinated multi-agent imitation
learning framework for player trajectory prediction in soccer, featuring
a training procedure that alternates between individual and team
policies. This ghosting approach later appears in basketball and
American football. \citet{seidl2018bhostgusters} develop a player
sketching system that generates basketball ghost defenders, while
\citet{schmid2021simulating} simulate defensive trajectories in the NFL
and perform evaluation using a completion probability model.
Furthermore, \citet{felsen2018where}, \citet{gu2023deep}, and
\citet{fassmeyer2025interactive} use deep latent variable models to
forecast multi-agent movements in team sports. \citet{groom2026machine}
assess off-ball defensive performance during corner kicks in soccer by
defining the so-called role-conditioned ghosts based on tactical and 
situational context.

Despite these contributions, this body of work still lacks proper
uncertainty consideration for the hypothetical player trajectories,
leaving room for further developments. Recently, \citet{yurko2026nfl}
present the first ghosting framework from a statistical viewpoint with
appropriate consideration that hypothetical players come from a
distribution. Using conditional density estimation, this work evaluates
defensive pass coverage by comparing the observed defender positioning
at \emph{a single moment} within a play (namely, when the receiver
catches the football) with a distribution of hypothetical defenders. The
hypothetical comparison is done using expected points---a commonly-used,
interpretable utility function for play valuation and in-game decision
making in the NFL \citep{romer2006firms, yurko2019nflwar}.
\citet{bajons2026pep} take a similar approach to evaluate tackling by
comparing a defender's value relative to a hypothetical missed tackle.

From here, a natural next step is to move beyond one single time point
and perform hypothetical evaluation at \emph{every moment} within a
play. To accomplish this, a reasonable strategy is to explicitly model
frame-level player movement and then design a simulation analysis to
generate player trajectories.
\citet{stokes2024generative}
propose Bayesian movement models using horse racing tracking data based on
forward and lateral distances, while controlling for
dynamic within-race features and random effects for race context,
jockeys, and horses. Once the models are fit, it is straightforward to
simulate a full race via posterior predictive simulation.
The simulation results can then be used to generate valuation measures such
as within-race finishing placement probabilities and provide insights
into different competition strategies.

\subsection{Summary of contributions}\label{summary-of-contributions}

%
%

In this work, we develop a novel generative modeling approach for
hypothetical evaluation of player positioning throughout a play in American football.
Using NFL tracking data, we focus on evaluating the observed tracking
information of an individual ball carrier in running plays. In our
framework, we characterize and model frame-level player movement based
on two attributes: step length (distance between consecutive locations)
and turn angle (angle formed by consecutive steps), inspired by the
animal movement literature. This approach offers several key advantages.
Our models include spatial features derived from the fine-grained
tracking data as well as player-specific random effects. This allows us
to gain practical insights into the movement profiles of NFL ball
carriers.
Moreover, since the models are fit in a Bayesian multilevel framework,
we leverage posterior predictive simulation to generate hypothetical
next steps for an individual ball carrier in each play with proper
uncertainty propagation. 
Here, the
observed step at each frame is compared to a posterior predictive set of
local alternative steps.
Thus, our simulations reflect
hypothetical individual movement behavior for a single step,
rather than providing a fully dynamic recreation of the actual play.

\begin{figure}[htbp]

{\centering \includegraphics[width=0.8\linewidth]{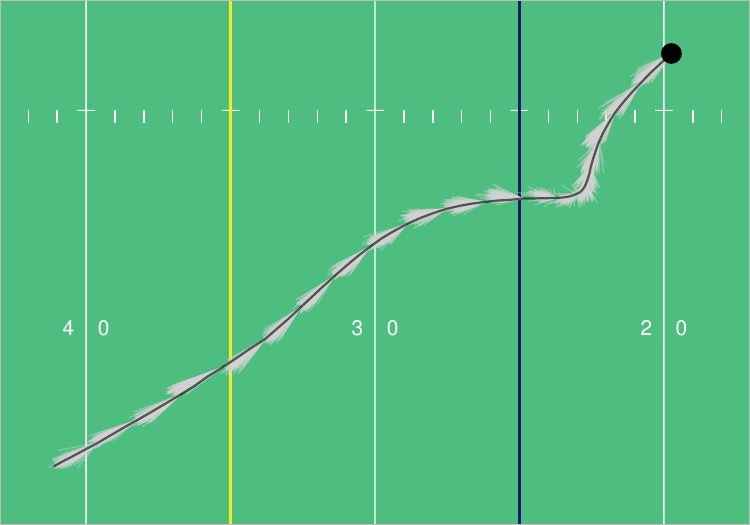} 

}

\caption{Simulated steps for an example play. The black path denotes the observed ball carrier trajectory after the handoff. At each frame along this path, a distribution of hypothetical next steps is generated (gray segments), against which the observed player movement can be evaluated. The blue and gold vertical lines represent the line of scrimmage and first down marker, respectively.}\label{fig:sim-traj}
\end{figure}

Figure \ref{fig:sim-traj} shows a visual illustration of our 
player evaluation strategy applied to
an example play. Here, the observed ball carrier trajectory is evaluated
at each frame against a local distribution of simulated steps. This ultimately enables player evaluation through hypothetical comparisons with a play value measure, which we illustrate using the expected ending yard line at different moments within a play.
Our approach provides a novel framework for evaluating
the observed frame-by-frame player tracking
information with proper consideration that the hypothetical movement
comes from a distribution. We note that our proposed methodology can be
extended to other settings in American football, as well as other team
sports such as basketball \citep{cervone2016multiresolution}, ice hockey
\citep{radke2023presenting} and soccer \citep{fernandez2021framework}.

The remainder of this manuscript is organized as follows. In Section
\ref{sec:data}, we discuss the NFL tracking data used in this study. 
In Section \ref{sec:methods}, we present our framework for evaluating player movement, including multilevel movement models,  simulation procedure and hypothetical evaluation strategy. Next, we analyze our results in Section
\ref{sec:results}, and provide our concluding remarks in Section
\ref{sec:discussion}.

\section{Data}
\label{sec:data}

In this work, we rely on player tracking data provided by the NFL Big
Data Bowl 2025 \citep{lopez2024nfl}, which span the first nine weeks of
the 2022 NFL season. This consists of two-dimensional coordinates for
all 22 players on the field and the football, recorded at 10 frames per
second. In addition, information about player movement such as speed,
acceleration, direction and orientation are included for each player at
each frame within a play. We also have access to different event tags
(e.g., ball snap, handoff, touchdown, etc.) for specific frames within
each play.

Table \ref{tab:tracking} shows a tracking data example for a run play by
running back Javonte Williams, which takes place in a week 2 game
between the Houston Texans and Denver Broncos during the 2022 NFL
regular season. For more context, Denver is on offense and starts the
play at their own 25-yard line. The play results in a 17-yard run by
Broncos running back Javonte Williams, who picks up a first down before
being tackled by the defense at the 42-yard line. In Section
\ref{sec:results}, we use this play as an example to illustrate our
proposed framework.

For our analysis, we focus on run plays by running backs in the NFL.
Typically, a run play starts with the quarterback taking the snap from
the center (\texttt{ball\_snap}) and handing the ball to a running back
(\texttt{handoff}). From here, the main goal of the running back is to
carry the ball and gain yards for the offensive team. In the example
play, the running back experiences first contact created by the defense
(\texttt{first\_contact}) before crossing the line of scrimmage and
first down marker. The running back continues to motion forward, and the
play eventually concludes with a tackle made by the defense
(\texttt{tackle}).

For each play, using the event labels in the tracking data, we extract
only frames within the so-called \emph{ball carrier sequence}.
Specifically, we identify the beginning of this time window as when the
handoff (from the quarterback to the running back) occurs, and the end
as any of the following outcomes: tackle, out-of-bounds, or touchdown.
After preprocessing, our final sample corresponds to 5,400 total plays
across all 136 games played during weeks 1 through 9 of the 2022 NFL
regular season.

\begin{table}[t]
\caption{Player tracking data for a play during the Houston Texans--Denver Broncos game during week 2 of the 2022 NFL season. The data shown here are for Broncos running back Javonte Williams, and the frames included are between the ball snap and tackle events. \label{tab:tracking}}
\centering
\begin{tabular}{rrrrrrrrl}
\hline
frameId & x & y & s & a & dis & o & dir & event \\
\hline
70 & 93.14 & 30.08 & 0.00 & 0.00 & 0.01 & 274.09 & 141.72 & \texttt{ball\_snap} \\
\vdots & \vdots & \vdots & \vdots & \vdots & \vdots & \vdots & \vdots & \vdots \\
84 & 91.30 & 27.79 & 5.09 & 2.54 & 0.50 & 249.50 & 219.68 & \texttt{handoff} \\
\vdots & \vdots & \vdots & \vdots & \vdots & \vdots & \vdots & \vdots & \vdots \\
110 & 85.52 & 19.04 & 3.84 & 3.15 & 0.37 & 276.28 & 270.68 & \texttt{first\_contact} \\
\vdots & \vdots & \vdots & \vdots & \vdots & \vdots & \vdots & \vdots & \vdots \\
149 & 68.63 & 5.11 & 3.37 & 4.11 & 0.30 & 316.78 & 234.09 & \texttt{tackle} \\
\hline
\end{tabular}
\end{table}

Using the provided tracking data, we construct features for our models in Section \ref{sec:methods}. To do so, we extract
spatial variables that reasonably summarize ball carrier's location and
trajectory, as well as their relationship with other players throughout
a play. We use an anchoring strategy as described in \citet{horton2020learning}
and \citet{yurko2020going} to derive the dynamic within-play features of
interest. In particular, we create features for three player groups:
ball carrier, offense (excluding ball carrier), and defense. Here, we
consider the ball carrier as the anchor point and order the offensive
and defensive players based on their Euclidean distance to the ball
carrier (e.g., defender 1 represents the nearest defender, and so on.) A
full list of our tracking data features is provided in Table
\ref{tab:features}.

\begin{table}[t]
\caption{List of features derived from player tracking data for three player groups: ball carrier, defense (11 players on each play), and offense (10 players on each play, excluding ball carrier). Note that horizontal and vertical directions are with respect to the end zone and sideline, respectively.}
\label{tab:features}
\centering
\begin{tabular}{p{0.6\textwidth}w{c}{0.125\textwidth}w{c}{0.08\textwidth}w{c}{0.08\textwidth}}
\hline 
\multirow{2}{*}{Feature} & \multicolumn{3}{c}{Player group} \\ 
& Ball carrier & Defense & Offense \\ 
\hline \addlinespace[0.5ex]
Horizontal yards from target endzone & \checkmark & \checkmark & \checkmark \\ \addlinespace[0.5ex]
Vertical yards from center of the field with respect to target endzone (positive: left side; negative: right side) & \checkmark & \checkmark & \checkmark \\ \addlinespace[0.5ex]
Horizontal yards from first down line & \checkmark & & \\ \addlinespace[0.5ex]
Speed (yards/second) & \checkmark & \checkmark &\checkmark \\ \addlinespace[0.5ex]
Angle of motion relative to ball carrier (radians) & & \checkmark & \checkmark \\ \addlinespace[0.5ex]
Horizontal yards relative to ball carrier & & \checkmark  & \checkmark \\ \addlinespace[0.5ex]
Vertical yards (absolute value) relative to ball carrier & & \checkmark  & \checkmark \\ \addlinespace[0.5ex]
Distance from ball carrier (yards) & & \checkmark  & \checkmark \\ \addlinespace[0.5ex]
\hline 
\end{tabular}
\end{table}

\section{Framework for evaluating player movement}
\label{sec:methods}

\subsection{Evaluating player movement}
\label{sec:eval-motivation}

Our framework is motivated by a general question in player evaluation from tracking data:
\emph{How valuable is the movement of a player relative to other
movements that could reasonably have been made in the same context?}
We evaluate frame-level player movement through a measure of instantaneous play value.
More generally, let $Z_{ijt}$ denote a generic play value quantity associated with the
movement of player $j$ at frame $t$ of play $i$. The specific definition of $Z_{ijt}$ depends on the application. For example, for a ball carrier in American football, play value
could be determined as a function of the end-of-play yard line, whereas for a receiver it could reflect the probability of completing a pass following their movement. The choice of play value measure is separate
from how player movement is characterized and modeled, allowing the same approach
to be applied to different settings and sports.

In this work, we focus on evaluating NFL ball carrier movement.
Let $L_{ijt}$ denote the end-of-play yard line from the current position on the field for ball carrier $j$ at frame $t$ of play $i$.
We can estimate
\begin{equation}
\label{eop-yardline}
\ell_{ijt} = E[L_{ijt} \mid \boldsymbol{X}_{ijt}],
\end{equation}
where $\boldsymbol{X}_{ijt}$ contains covariate information about the locations and trajectories of all players up to frame $t$.

The predicted outcome associated with the observed state ($\hat \ell_{ijt}$), however, does not by itself indicate how effective the player movement is.
Instead, we need to evaluate the observed movement relative to
other alternatives available at that moment given the same context.
Thus, we consider player movement evaluation through hypothetical comparison.

For each hypothetical movement $h$, consider the corresponding predicted end-of-play yard line $\hat \ell_{ijt}^{(h)}$, for $h = 1, \dots, H$. The resulting collection $\{\hat \ell_{ijt}^{(1)}, \dots, \hat \ell_{ijt}^{(H)}\}$ provides a distribution of predicted outcomes associated with hypothetical movements.
This distribution allows us to evaluate the observed movement relative to feasible alternatives.

The statistical challenge here is to construct the
distribution of hypothetical movements.
At any given moment within a play, we observe only one realized movement
among the many movements that could have occurred under the same context.
As such, generating meaningful hypothetical comparison requires explicitly modeling frame-level player movement to characterize how players move conditional on the surrounding environment.
Ultimately, this can be used to obtain simulated movements and, consequently, a distribution of play values against
which the observed movement can be evaluated.

This motivates the methodological formulation of our framework. 
Below, we first specify a representation of frame-level player movement, then carefully build statistical models for its constituent movement characteristics, and use the fitted models to generate hypothetical movements for evaluation.
Although we develop and illustrate our approach for NFL ball carriers, it can be adapted to other player roles, sports, and tracking settings.

\subsection{Characterizing player movement}\label{sec:characterization}

Our first task is to characterize the movement of a player at every
moment within a play. We aim to find an appropriate representation by decomposing the total movement into components
that are features derived from tracking data.
In American football, due to the complex nature of many
athletes interacting on the field, player movement consists of multiple elements: bursts of speed, changes in direction, and lateral adjustments, to name a few. This necessitates a formulation that captures notions of both \emph{displacement} and
\emph{directionality}, in order to properly describe frame-level
movement.



\begin{figure}[t]

{\centering \includegraphics[width=0.45\linewidth]{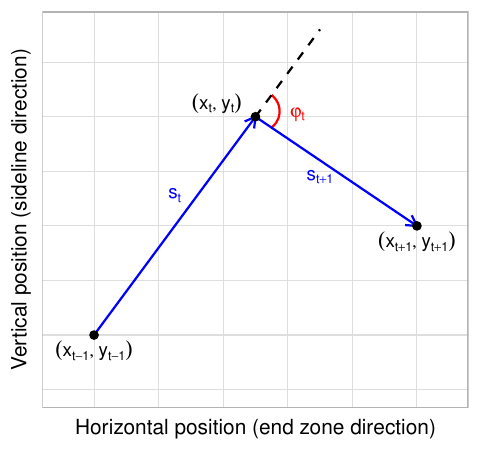} 

}

\caption{Illustration of the step-and-turn characterization for player movement. The coordinate axes correspond to standardized football field coordinates, where the horizontal axis represents the end zone direction and the vertical axis represents the sideline direction.}\label{fig:step_turn_illustration}
\end{figure}

We therefore use a step-and-turn representation to characterize player movement. 
Specifically, for a player at each time point within a play, their movement is
composed of two attributes: \emph{step length} (i.e., the distance
between successive locations) and \emph{turn angle} (i.e., the angle
between successive displacement vectors). In Figure
\ref{fig:step_turn_illustration}, we present a visual explanation of the
two aforementioned movement characteristics.

Formally, let
\((x_t, y_t)\) be the observed player location at frame \(t\) of a
play, for \(t=1, \dots, T + 1\). Then, for an individual player at frame
\(t\):

\begin{itemize}
\item
  The step length \(s_t\) is the Euclidean distance between two points
  \((x_t, y_t)\) and \((x_{t+1}, y_{t+1})\), expressed as
  \(s_{t} = \sqrt{(x_{t+1} - x_t)^2 + (y_{t+1} - y_t)^2}\).
\item
  The turn angle \(\varphi_t\) is the change in bearing between two
  consecutive time intervals \([t-1, t]\) and \([t, t + 1]\). That is,
  \(\varphi_t = b_t - b_{t-1}\), where the bearing angle
  \(b_t \in [-\pi, \pi]\) is defined as
  \(b_t = \text{atan}2 (y_{t+1} - y_t, x_{t+1} - x_t)\).
\end{itemize}

As a side note, observant readers will recognize the connection between
our choice of player movement metrics and the step-and-turn
characterization in the animal movement literature. Indeed, the step
length and turn angle computed from telemetry data are vital components
in models for understanding the movement and habitat preferences of
animals. For complete surveys on statistical methods for animal tracking
data, see \citet{hooten2017animal} and \citet{leosbarajas2026}.

\begin{figure}[t]

{\centering \includegraphics[width=0.65\linewidth]{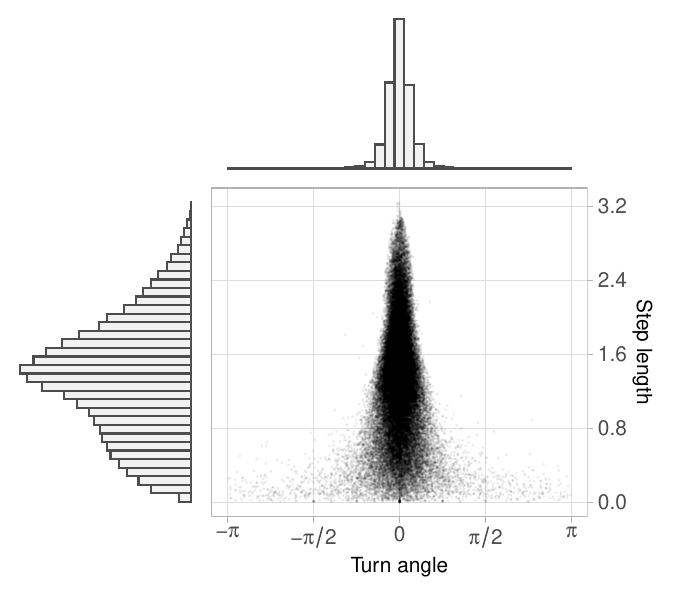} 

}

\caption{Joint and marginal distributions of step length and turn angle for running backs on run plays during the first nine weeks of the 2022 NFL season.}\label{fig:step_turn_distributions}
\end{figure}

For context, Figure \ref{fig:step_turn_distributions} shows the joint
and marginal distributions of the two movement quantities for NFL
running backs during the first nine weeks of the 2022 season. We notice
a somewhat asymmetric, bimodal step length distribution, while the turn
angle values are highly concentrated around zero. We also observe that
variability in turn angle decreases as step length increases. For small
steps, turn angles are widely dispersed across the full range of value,
indicating more erratic movement directions. As the step gets larger,
the spread of turn angles narrows sharply around zero, meaning longer
steps are typically taken in straighter directions with little angular
deviation. This suggests that players tend to make more variable turns
during short movements, but move more consistently when taking longer
strides.

From here, our model choice focuses on five qualities: (i) appropriate
response distributions, (ii) dependence between movement outcomes, (iii)
spatial tracking data covariates, (iv) player random effects, and (v)
generative capability. With these aspects in mind, we develop our models
for step length and turn angle as follows.

\subsection{Step length model}\label{sec:step-model}

Let \(s_{ijt}\) denote the step length for ball carrier \(j\) at frame
\(t\) of play \(i\), and \(\tilde s_{ijt}\) denote its transformed
version (via a scaled arcsin transformation; see Appendix \ref{sec:step-assess} for our justification and model assessments.)
We fit the following multilevel model for \(\tilde s_{ijt}\):
\begin{equation}
\label{eq:step}
\begin{aligned}
\tilde s_{ijt} &\sim \mathcal N(\mu_{ijt}^{\textsf{(SL)}}, \sigma^2),\\
\mu_{ijt}^{\textsf{(SL)}} &= \alpha_0^{\textsf{(SL)}} + \boldsymbol X^{\textsf{(SL)}}_{ijt} \boldsymbol{\beta^{\textsf{(SL)}}}  + u_j + v_k,\\
u_j &\sim \mathcal N(0, \tau^2_u),\\
v_k &\sim \mathcal N(0, \tau^2_v).
\end{aligned}
\end{equation} In this model, we assume a Gaussian distribution for the
response \(\tilde s_{ijt}\). We include random intercepts \(u_j\) and
\(v_k\) for two groups: ball carrier \(j\) and defensive team \(k\),
respectively. We also account for tracking data features as described in
Section \ref{sec:data} through \(\boldsymbol X^{\textsf{(SL)}}_{ijt}\)
and estimate the coefficients \(\boldsymbol{\beta^{\textsf{(SL)}}}\) as
fixed effects. Rather than using features for all players, we choose a
simpler set of covariates for ease of model fitting, as these features
are treated as linear fixed effects in our model. Specifically, we
control for frame-level features describing the ball carrier and the
closest defender at each frame (see Table \ref{tab:features}). In addition, we account for the number
of defenders and offensive teammates in each direction (left/right and
front/back) of the ball carrier.
Importantly, we condition on the step length from the previous frame as
part of our covariates \(\boldsymbol X^{\textsf{(SL)}}_{ijt}\). This
captures the fact that a ball carrier's distance traveled over
successive time points are likely to be correlated. In connection with
speed, this makes intuitive sense: if a running back is covering
distance at a certain rate, they are also likely to maintain a similar
speed in the next time step, leading to correlation over time.

\subsection{Turn angle model}\label{sec:turn-model}

Let \(\varphi_{ijt}\) denote the turn angle for ball carrier \(j\) at
frame \(t\) of play \(i\). We fit the following multilevel model for \(\varphi_{ijt}\):
\begin{equation}
\label{eq:turn}
\begin{aligned}
\varphi_{ijt} &\sim \textsf{von Mises}(\mu_{ijt}^{\textsf{(TA)}}, \kappa_{ijt}),\\
\tan \frac{\mu_{ijt}^{\textsf{(TA)}}}{2} &= \alpha_0^{\textsf{(TA)}} + \boldsymbol X^{\textsf{(TA)}}_{ijt} \boldsymbol{\beta^{\textsf{(TA)}}},\\
\log \kappa_{ijt} &= \gamma_0 + \gamma_1 s_{ijt} + w_j,\\
w_j &\sim \mathcal N(0, \tau^2_w).
\end{aligned}
\end{equation} Here, we use a von Mises response distribution for the
instantaneous turn angle \(\varphi_{ijt}\) and explicitly both the mean
\(\mu_{ijt}^{\textsf{(TA)}}\) and concentration \(\kappa_{ijt}\)
parameters. To model \(\mu_{ijt}^{\textsf{(TA)}}\), we use a tan-half
link function and include a simple set of tracking data features similar
to the step length model in (\ref{eq:step}) (i.e., features for ball
carrier, closest defenders, and number of players in each direction).
The covariate information \(\boldsymbol X^{\textsf{(TA)}}_{ijt}\) also
includes the turn angle from the previous frame \(\varphi_{ij,t-1}\).
This captures the notion of directional persistence, reflecting a
player's tendency to make consecutive turns in a similar direction.

Next, to model the concentration parameter \(\kappa_{ijt}\), we use a
log link function and condition on the step length at the current frame
\(s_{ijt}\). This effectively captures the dependence between the step
and turn movement characteristics. We believe it is more natural to
condition on step length in the turn angle model rather than the other
way around. Specifically, the step length dictates how much distance is
covered in a single movement, and turning behavior varies with how far a
player has moved (see Figure \ref{fig:step_turn_distributions}). Once
the step is established, the turn angle determines the directional
change relative to the previous step. If the step length is small, both
small and large turns can be executed, whereas longer steps make it more
difficult to perform sharper turns.

Further, we include a random intercept \(w_j\) for ball carrier \(j\)
when modeling \(\kappa_{ijt}\). This enables us to estimate the
differences in turn angle variability among NFL running backs. Furthermore, we can infer a player's
ability to display variable change of direction, which is a legitimate athletic trait from a
player evaluation point of view.


\subsection{Model fitting}\label{sec:fitting}

We fit both multilevel models (\ref{eq:step}) and (\ref{eq:turn}) for step length
and turn angle using a Bayesian approach with \texttt{Stan}
\citep{carpenter2017stan} via the \texttt{brms} package in \texttt{R}
\citep{burkner2017brms, r2025language}. This provides natural
uncertainty quantification for all model parameters, whose posterior
distributions are estimated using Markov chain Monte Carlo (MCMC)
through a no-U-turn sampler \citep{hoffman2014nuts}. We use weakly
informative priors for the parameters in both step length and turn angle
models. Specifically, for the standard deviation parameters (\(\sigma\),
\(\tau_u\), \(\tau_v\), and \(\tau_w\)), we choose vague
\(\text{half-}t_3\) priors \citep{gelman2006prior}. For the remaining
model parameters, we specify default prior distributions provided by
\texttt{brms}. Specifically, we assume
\(\alpha_0^{\textsf{(SL)}} \sim t_3\),
\(\alpha_0^{\textsf{(TA)}} \sim t_1\),
\(\gamma_0 \sim \mathcal N(5, 0.8^2)\), and uniform priors for all fixed
effect coefficients \(\boldsymbol{\beta^{\textsf{(SL)}}}\) and
\(\boldsymbol{\beta^{\textsf{(TA)}}}\).
In Appendix \ref{sensitivity}, we provide a sensitivity analysis of different prior specifications for the variance parameters in our models.
We find that the posterior estimates of the variance components and player random effects are highly consistent, indicating that our results are robust across different prior choices.

For both models (\ref{eq:step}) and (\ref{eq:turn}), we use 4 parallel chains, each with 5,000 iterations and a warm-up of 2,500 draws.
As a result, we base our inference on 10,000 posterior samples for each model in the forthcoming analysis. We observe
evidence of convergence of the MCMC sampling procedure according to
trace plots and \(\hat R\) values close to 1
\citep{gelman1992inference}. We also observe no problematic effective
sample size for each parameter \citep{gelman2013bayesian}.
All MCMC diagnostics are shown in Appendix \ref{mcmc-convergence}.

After model fitting, when of interest, we sample from the posterior
predictive distributions \begin{equation}
\label{eq:step-ppd}
\int_{\Theta_{\tilde s}} p(\tilde s_{\textsf{new}} \mid \theta_{\tilde s}) \pi (\theta_{\tilde s} \mid \mathcal D_{\tilde s}) d \theta_{\tilde s} \quad \text{ and } \quad
\int_{\Theta_{\varphi}} p(\varphi_{\textsf{new}} \mid \theta_{\varphi}) \pi (\theta_{\varphi} \mid \mathcal D_{\varphi}) d \theta_{\varphi}
\end{equation} for step length and turn angle, respectively, where
\(\theta_{\tilde s}\) and \(\theta_{ \varphi}\) represent all model
parameters, and \(\mathcal D_{\tilde s}\) and \(\mathcal D_{\varphi}\)
denote all data.

\subsection{Simulating player movement}\label{sec:simulating}

Using our proposed Bayesian multilevel models (\ref{eq:step})
and (\ref{eq:turn}), we can
take advantage of posterior predictive simulation to generate
hypothetical player movement in any given play.
We focus on generating a distribution of possible next steps
for the observed player, predicting one step ahead at each frame.
In other words, this strategy simulates only the immediate next
movement, rather than generating full trajectories over multiple frames.
This distribution represents the set of local hypothetical steps that
could reasonably have been taken under the same on-field spatial
environment as the observed movement. Note that we only simulate
the individual player movement while holding the observed
tracking information of all other players fixed. As such, the
resulting simulations represent hypothetical player behavior
instead of a fully multi-agent replay of the actions within a play.
In Section \ref{sec:discussion}, we discuss challenges in extending
our framework to an iterative, frame-by-frame forward trajectory
simulation setting, which involves modeling the movement of multiple players on the field.


Our full simulation procedure is as follows. For a player at frame
\(t\) with observed location \((x_t, y_t)\) and bearing angle \(b_t\),
we use the posterior predictive distributions in (\ref{eq:step-ppd})
to simulate step length and turn angle for the next
frame \(t+1\). We generate \(H\) hypothetical next steps for a player at
frame \(t\) by repeating the following procedure for each draw
\(h = 1, \dots, H\):

\begin{enumerate}
\def\labelenumi{\arabic{enumi}.}
\item
  \textbf{Simulate step length.} First, we draw parameter values
  \(\theta_{\tilde s}^{(h)}\) from the posterior distribution
  \(\pi(\theta_{\tilde s} \mid \mathcal D_{\tilde s})\). To simulate a
  generic player, we draw a new random effect
  \(u_{\textsf{new}}^{(h)} \sim \mathcal{N}(0, \tau_u^{2^{(h)}})\),
  where \(\tau_u^{2^{(h)}}\) is the \(h^{\text{th}}\) posterior draw of
  the variance parameter. Since the random intercept represents 
  player-specific deviation from the population mean, drawing it from
  a distribution centered at zero produces the movement behavior of an average player.
  We then draw \(\tilde{s}_{t+1}^{(h)}\) from
  the sampling distribution
  \(p(\tilde{s}_{t+1} \mid \theta_{\tilde{s}}^{(h)})\) and
  back-transform to obtain the predictive step length \(s_{t+1}^{(h)}\).
\item
  \textbf{Simulate turn angle (conditional on step length).} Because the
  turn angle model controls for step length, we incorporate the step
  length predictions from Step 1. For each predictive step length draw
  \(s^{(h)}_{t+1}\), we draw one single conditional turn angle
  \(\varphi^{(h)}_{t+1}\). To do so, we first draw parameter values
  \(\theta_{\varphi}^{(h)}\) from the posterior distribution
  \(\pi(\theta_{\varphi} \mid \mathcal D_{\varphi})\). We then draw a
  new player random effect
  \(w_{\textsf{new}}^{(h)} \sim \mathcal{N}(0, \tau_w^{2^{(h)}})\).
  Finally, we draw a conditional turn angle \(\varphi_{t+1}^{(h)} \sim
  p(\varphi_{t+1} \mid s_{t+1}^{(h)}, \theta_{\varphi}^{(h)})\).
\item
  \textbf{Obtain hypothetical location and features.} Each pair
  \((s_{t+1}^{(h)}, \varphi_{t+1}^{(h)})\) defines a single hypothetical
  next step, which is then converted to a hypothetical next location
  \[(x_{t+1}^{(h)}, y_{t+1}^{(h)}) = \left( x_t + s_{t+1}^{(h)} \cos(\varphi_{t+1}^{(h)} + b_t), \; y_t + s_{t+1}^{(h)} \sin(\varphi_{t+1}^{(h)} + b_t) \right).\]
  We also compute relevant tracking data features
  \(\boldsymbol X^{(h)}_{t+1}\) for subsequent hypothetical evaluation.
\end{enumerate}

As established, the posterior predictive steps and turns represent
realizations of a generic, average player's movement. In Section
\ref{sec:results}, we present evaluation analyses using this
conventional representation of the hypothetical baseline. We emphasize
that our simulation framework is modular, as the hypothetical comparison
is fully customizable. By modifying the random effect specification, we
can substitute a different player identity for simulation, such as a
specific backup running back in our ball carrier application. It is also possible to assess a player against a distribution of simulated alternatives based on their
own movement profile using their individual-specific random effect,
and we present results for this setting in Appendix \ref{alt-sim}.

\subsection{Hypothetical evaluation}\label{sec:hypothetical-eval}

Based on the simulation results, we perform evaluation by comparing the observed
player positioning at different moments within a play with a distribution of hypothetical alternatives.
Here, we illustrate with the ball carrier application as mentioned in Section \ref{sec:eval-motivation}.

To estimate the end-of-play yard line $\ell_{ijt}$ in (\ref{eop-yardline}), we use a multinomial classification approach.
We consider the response at one-yard increments
corresponding to possible ending yard lines between 10 and 110, where 10 represents the offense's own end zone and 110 represents the target end zone. The model estimates the conditional probability of each ending location given the observed tracking features, capturing uncertainty over the range of possible yard lines.
We then obtain the expected ending yard line by summing over the possible yard lines weighted by their predicted probabilities.
That is,
\begin{equation}
\ell_{ijt}
=
E\left[
L_{ijt}
\mid
\boldsymbol{X}_{ijt}
\right]
=
\sum_{l=10}^{110}
l\,
P\left(
L_{ijt}=l
\mid
\boldsymbol{X}_{ijt}
\right).    
\end{equation}

We train a multinomial classification model to estimate  \(\ell_{ijt}\) using \texttt{CatBoost}, a gradient boosting-based library \citep{prokhorenkova2018catboost}.
All details regarding our end-of-play yard line model are provided in Appendix \ref{yards-model}.
Since our player evaluation framework is modular,
this model can be replaced by any other method for predicting the
ending yard line. For further discussion of this prediction problem, we point the reader to prior
work such as \citet{yurko2020going} and \citet{gordeev2020first}\footnote{We note that the task of modeling rushing yards
  is not the primary focus of our paper. Besides, we suspect that many
  NFL teams and vendors have their own version of the end-of-play yard line
  model, as this was a foundational development in football analytics
  when tracking data first became available.}.

We then perform hypothetical
evaluation for a ball carrier at a given frame within a play as follows.

\begin{itemize}
\item
  \textbf{Estimate ball carrier ending yard line.} First, we obtain \(\hat \ell_{ijt}\) for player \(j\) 
  given their observed location at frame \(t\) during play \(i\).
\item
  \textbf{Estimate hypothetical ending yard line.} Next, we obtain the
  predicted values
  \(\hat \ell_{ijt} ^{(1)}, \dots, \hat \ell_{ijt} ^{(H)}\) for each
  location associated with hypothetical step \(h=1, \dots, H\) at frame \(t\) during play
  \(i\). This gives us a distribution of
  hypothetical ending yard line across different simulated steps
  at a given moment in a play.
\item
  \textbf{Evaluate ball carrier movement.} We then compute the
  difference \(\delta_{ijt}\) in the predicted ending yard line between the
  location of ball carrier \(j\) and each hypothetical step
  \(h=1, \dots, H\) at frame \(t\) during play \(i\). That is,
  \begin{equation}
  \label{eq:delta}
  \delta_{ijt} ^{(h)} = \hat \ell_{ijt} - \hat \ell_{ijt} ^{(h)}.
  \end{equation}
\end{itemize}

Ultimately, this allows us to compare the observed ball carrier movement
against a distribution of simulated alternatives in terms of a
play outcome quantity (i.e., yards). For instance, at each time
point \(t\) of play \(i\), we can average across the distribution of
hypothetical steps \(h=1, \dots, H\) to obtain a point estimate
\begin{equation}
\label{eq:delta-average}
\bar \delta_{ijt} = \frac{1}{H}\sum_{h=1}^H \delta_{ijt} ^{(h)}
\end{equation} for the yardage difference between ball carrier
\(j\) and the simulated baseline, along with corresponding
quantile-based interval estimates. We can then summarize across
different moments within each play and produce a variety of player
valuation metrics, which is illustrated in Section \ref{metrics}.
We believe this presents a practical approach for summarizing the
complexity of player tracking data into intuitive measures of player
performance.

\section{Results}
\label{sec:results}

\subsection{Inference on model parameters: player
ratings}\label{inference-on-model-parameters-player-ratings}

\begin{figure}[!htbp]

{\centering \includegraphics[width=0.6\linewidth]{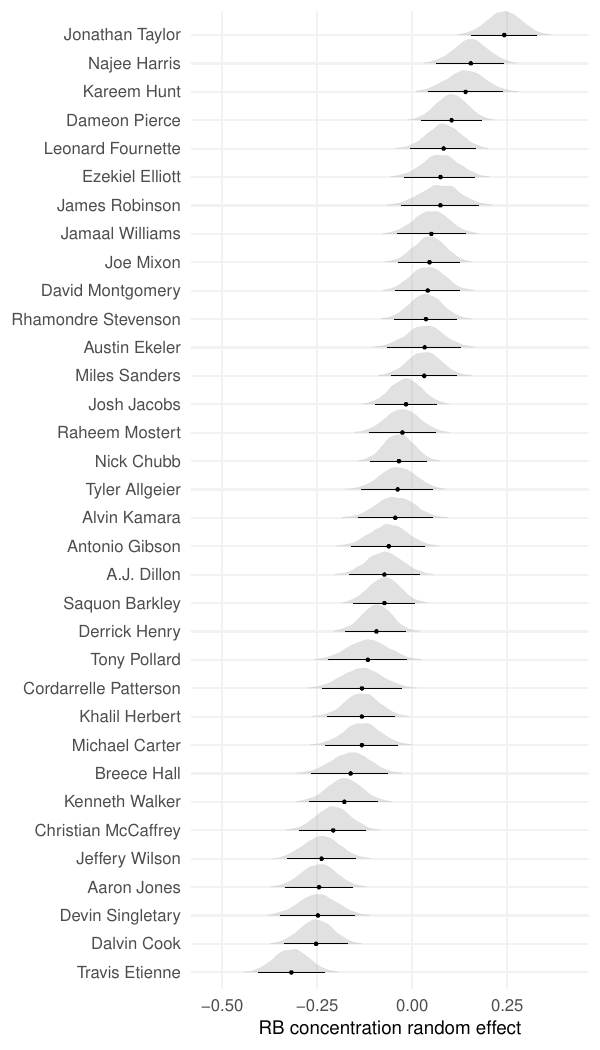} 

}

\caption{Posterior distributions of the turn angle concentration random effect $w_j$ for NFL running backs with at least 70 rush attempts on running plays over the first nine weeks of the 2022 regular season. For each player, the posterior mean and corresponding 95\% credible interval are depicted.}\label{fig:angle_posterior_distributions}
\end{figure}

Our multilevel models allow us to gain insights into the step-and-turn 
movement of NFL running backs during the first nine weeks of the 2022 season.
Figure \ref{fig:angle_posterior_distributions} shows posterior
distributions of the concentration random effect \(w_j\) in the turn
angle model for running backs with at least 70 rush attempts across the
considered plays. Here, the players are arranged by their posterior mean
estimates, which demonstrate the ability to display variable change of
direction. In particular, a higher posterior mean is associated with
lower variability (or greater concentration) in making directional
changes when carrying the ball.

In this leaderboard, we clearly see Jonathan Taylor as the player with
the least variability in turn angle relative to the rest of the ball
carriers. This considerably makes sense, as Taylor is known to have a
tendency to run straight ahead with speed rather than making variable
directional adjustments (e.g., making cuts). In addition, the scouting
report from his college career indicates that ``{[}h{]}e's more of a
straight-ahead runner than make-you-miss guy\ldots{}''
\citep{kramer2019jonathan}. This further validates our results and lends
more credibility to our estimates as an effective measure of ball
carrier's change of direction variability.

Note that there are substantial differences in the credible intervals
among this subset of running backs. In particular, there are no overlaps
between the intervals between the top and bottom ball carriers in this
leaderboard. Thus, our estimates provide discriminative power for
differentiating between the players, which is a useful statistical
property of a performance metric in sports \citep{franks2016meta}.

\begin{figure}[htbp]

{\centering \includegraphics[width=0.6\linewidth]{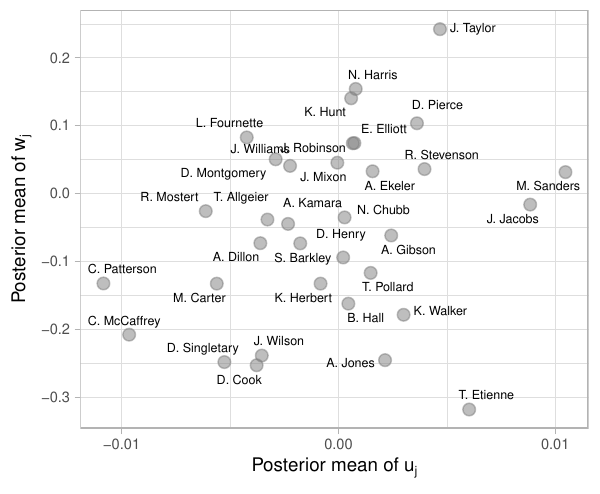} 

}

\caption{Relationship between the posterior means of the running back random effects $u_j$ (on a transformed scale) and $w_j$ when modeling the mean step length and turn angle concentration, respectively $(r=0.279)$.  Results shown here are for NFL running backs with at least 70 rush attempts on running plays over the first nine weeks of the 2022 regular season.}\label{fig:step_angle_posterior_means}
\end{figure}

Next, we examine the running back random effect estimates in both step
length and turn angle models together, to gain more understanding of the
player movement profiles. For the same subset of players as before,
Figure \ref{fig:step_angle_posterior_means} shows the joint distribution
for the posterior means of the running back random effects \(u_j\) and
\(w_j\) in the step length and turn angle models, respectively (see also
Appendix \ref{step-ratings} for our step length leaderboard).
We notice different traits of running backs as revealed by the
scatterplot. For example, Jonathan Taylor displays long strides but low
variability in turn angle when carrying the ball. As alluded to earlier,
Taylor tends to lean on speed in his running game more so than making
directional adjustments. On the other hand, Christian McCaffrey is a
shifty ball carrier that covers shorter distance between time steps.
This may indicate that McCaffrey exhibits lateral agility and can change
direction rapidly, while also is a patient runner that relies on shorter
strides to stay balanced. Overall, our results suggest that some players
are more effective at running straight with speed to gain ground
quickly, while others excel at making erratic, variable turns.

\subsection{Example play evaluation}\label{sec:example-play}

To illustrate, we use the example play mentioned in Section
\ref{sec:data} and assess the observed movement of Denver Broncos
running back Javonte Williams throughout this play. Figure
\ref{fig:sim-traj} shows the actual trajectory for Williams, as well as
100 simulated steps at each time point along this path, following our
simulation procedure in Section \ref{sec:simulating}. Using the
distribution of hypothetical steps at each frame, we can obtain
relevant summary about the play.

\begin{figure}[!htbp]

{\centering \includegraphics[width=0.7\linewidth]{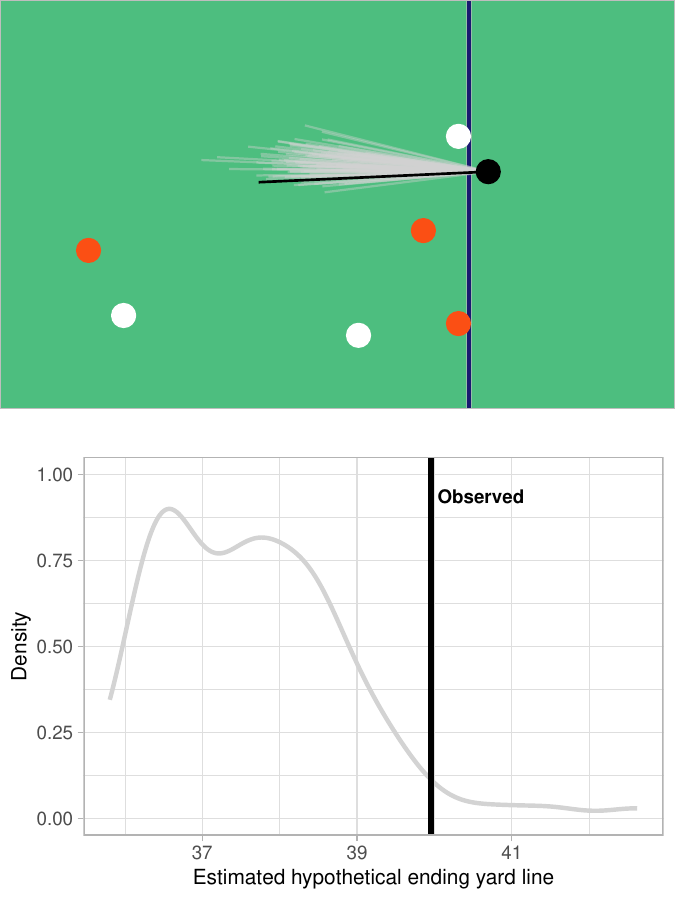} 

}

\caption{\textit{Top:} Simulated ball carrier steps for the moment of first contact in the example play. The offense's moving direction is from right to left. The black line represents the observed ball carrier step for this frame, and the gray lines represent the hypothetical steps obtained from 100 simulations. The blue vertical line denotes the line of scrimmage, and other offensive and defensive players nearby are highlighted in orange and white, respectively. \textit{Bottom:} Distribution of hypothetical ending yard line compared to the estimated ending yard line by ball carrier at first contact.}\label{fig:first-contact-steps}
\end{figure}

We now focus on a single snapshot at the first contact event to
illustrate our step selection analysis. Figure
\ref{fig:first-contact-steps} shows the simulated steps for the moment
of first contact, demonstrating how our generative approach captures the
variability and alternative possibilities of player movement. We also
display the estimated ending yard line for the ball carrier compared to the
distribution of hypothetical ending yard line. We see that the observed
performance estimate lies in the upper tail (above the
95th percentile) of the hypothetical distribution. Thus, the actual
player positioning yields a more favorable outcome than most of the
simulated alternatives at this particular instant of first contact.
We note that the same snapshot-level analysis can be done for other
moments during this play as well as other plays in our sample.

\begin{figure}[t]

{\centering \includegraphics[width=0.7\linewidth]{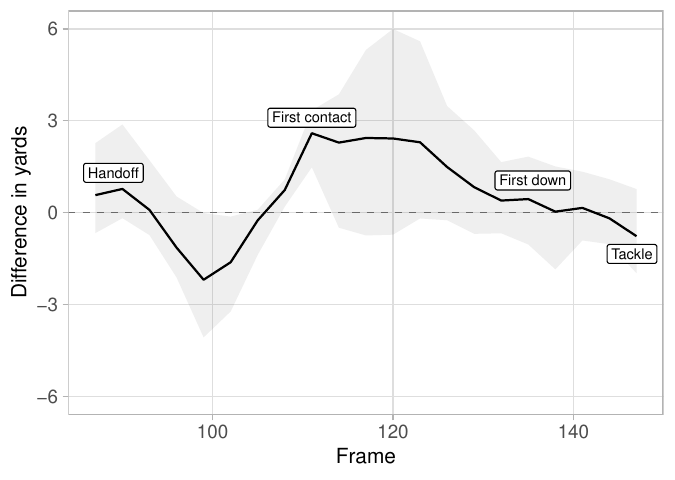} 

}

\caption{Changes in the difference in yards (with 95\% credible interval) between the observed and hypothetical ball carrier movement throughout the example play.}\label{fig:sim-deviation}
\end{figure}

Next, we compute the \(\bar \delta\) value for each frame as described
in Equation (\ref{eq:delta-average}) to evaluate the ball
carrier movement during this example play. Figure
\ref{fig:sim-deviation} shows the changes in \(\bar \delta\) as the play
progresses, which represents the average yardage difference
between the observed and hypothetical movement. We can integrate across
all moments in time to get an accumulated \(\bar \delta\) value for this
play. Overall, the observed ball carrier movement is worth +11.4
yards, relative to the hypothetical baseline. This aggregate can be
further broken down by smaller time segments within the play to identify
which phases positively or negatively contribute to the overall player
performance. This type of micro-level breakdown can provide actionable
insights for improving specific stages of the run.

Following the handoff, we observe a segment of frames with negative
difference in yards on average between the observed and hypothetical
player movement. This suggests early movement inefficiency by Javonte Williams,
who initially runs into a congested area in the actual play. However,
Williams manages to get out of traffic and begins to outperform the
simulated alternatives in later moments of this play. At the event of first
contact and adjacent frames, the yardage differential increases and
reaches peak performance at near +3 yards. This indicates a
positive deviation from the hypothetical baseline; thus post-contact
movement is highly efficient during this time window. In the actual
play, Williams bounces off multiple tackles during this phase of the
run. Noticeably, the interval band widens as the play develops,
reflecting increased variability and uncertainty in post-contact
dynamics. After the first down is secured, the yardage differential is near zero and eventually drops into negative values. Hence,
during the final phase of the run, the observed movement contributes
marginal value compared to the hypothetical alternatives.

\subsection{Player performance
metrics}\label{metrics}

For each of the 5,400 considered run plays from the first nine weeks of
the 2022 season, we implement the simulation procedure presented in
Section \ref{sec:simulating} and generate a distribution of 100
hypothetical ball carrier steps at each frame within a play. This
enables us to compare the observed and simulated player movement at any moment
in any given play. Our results can be further aggregated to construct
meaningful metrics for player evaluation. Here, we present two examples
of such measures: yards success rate and explosiveness.

\noindent \textbf{Example 1} (Yards success rate). We compute the
fraction of frames in which the \(\delta_{ijt}\) value defined in
(\ref{eq:delta}) exceeds zero. This indicates how often the observed
ball carrier performs better than the simulated alternatives in terms of
yards. We calculate this quantity as follows. Using the
distribution of \(h=1,\dots,H\) hypothetical steps generated for ball
carrier \(j\) at frame \(t\) of play \(i\), we compute \begin{equation}
\frac{1}{H} \sum_{h=1}^{H} \mathbbm{1}(\delta_{ijt}^{(h)} > 0).
\end{equation} This gives us a frame-level summary for the proportion 
of simulations where the actual positioning yields higher
estimated yardage than the hypothetical movement.
From here, we summarize the overall tendency of a ball carrier 
to outperform the hypothetical baseline by averaging over
all observed frames.

Table \ref{tab:fraction} provides a list of the top and bottom 5 players
based on the average fraction of positive yards relative to
hypothetical behavior. This captures how consistently a player exceeds the play value
they hypothetically would have gained. We notice
players at the top led by Josh Jacobs with repeatable ability to
generate positive yardage relative to the simulated baseline. On the
other hand, we observe less frequent and reliable yardage deviation for
ball carriers like Michael Carter and A.J. Dillon. This could be
indicative of inefficiencies in either individual running style or
offensive execution.

\begin{table}[t]
\caption{Leaderboard of NFL running backs (with at least 70 rush attempts over the first nine weeks of the 2022 regular season) according to our yards success rate measure on average via hypothetical comparison. \label{tab:fraction}}
\centering
\begin{tabular}{rllrr}
\hline
Rank & Player & Team & Plays & Yards success rate \\
\hline
1 & Josh Jacobs & LV & 134 & 0.542 \\
2 & Miles Sanders & PHI & 127 & 0.539 \\
3 & Travis Etienne & JAX & 106 & 0.527 \\
4 & Dameon Pierce & HOU & 135 & 0.512\\
5 & James Robinson & JAX/NYJ & 87 & 0.507 \\
\vdots & \vdots & \vdots & \vdots & \vdots \\
30 & Christian McCaffrey & CAR/SF & 104 & 0.457\\
31 & Raheem Mostert & MIA & 104 & 0.453 \\
32 & Cordarrelle Patterson & ATL & 70 & 0.450\\
33 & A.J. Dillon & GB & 95 & 0.448\\
34 & Michael Carter & NYJ & 81 & 0.446 \\
\hline
\end{tabular}
\end{table}

\noindent \textbf{Example 2} (Explosiveness). We quantify a ball
carrier's tendency to produce outcomes in the extreme right tail of the
hypothetical ending yard line distribution. Specifically, we measure whether
the observed yardage at each frame exceeds a high quantile of the
posterior predictive distribution of hypothetical play values. To some
extent, this captures a player's ability to generate rare, high-impact
plays and extreme outcomes (e.g., long gains, breakaway touchdowns,
etc.)

For a ball carrier \(j\) at frame \(t\) during play \(i\), we compute
\begin{equation}
\mathbbm{1}(\hat \ell_{ijt} > q_{0.95}(\hat \ell^{(h)}_{ijt})),
\end{equation} where \(q_{0.95}(\hat \ell^{(h)}_{ijt})\) denotes the
95\% quantile of the hypothetical ending yard line distribution at frame \(t\),
for \(h=1,\dots,H\). We then average over all frames within
a play, and across all observed plays for each ball carrier to obtain
a player-level summary of explosiveness.

Table \ref{tab:explosiveness} reports a leaderboard of NFL running backs rated by the proposed
explosiveness measure. Our metric identifies players such as Travis
Etienne and Kenneth Walker as the leaders in the rankings. These running
backs are widely recognized for their explosive burst and ability to
break off long-yardage plays. In contrast, we observe lower
explosiveness values for players at the bottom of our list---only around
half those of the top performers. In particular, A.J. Dillon, Jamaal Williams and Tyler
Allgeier are the least likely to produce extreme yardage outcomes among
the considered running backs. This aligns with their rushing profiles,
which are typically characterized by powerful, short gains in short
yardage situations rather than frequent explosive plays.

\begin{table}[t]
\caption{Leaderboard of NFL running backs (with at least 70 rush attempts over the first nine weeks of the 2022 regular season) according to our explosiveness measure on average via hypothetical comparison. \label{tab:explosiveness}}
\centering
\begin{tabular}{rllrr}
\hline
Rank & Player & Team & Plays & Explosiveness \\
\hline
1 & Travis Etienne & JAX & 106 & 0.118 \\
2 & Aaron Jones & GB & 100 & 0.095 \\
3 & Kenneth Walker & SEA & 94 & 0.093 \\
4 & Miles Sanders & PHI & 127 & 0.092 \\ 
5 & Antonio Gibson & WAS & 86 & 0.089 \\
\vdots & \vdots & \vdots & \vdots & \vdots \\
30 & Michael Carter & NYJ & 81 & 0.054 \\
31 & David Montgomery & CHI & 103 & 0.054 \\
32 & A.J. Dillon & GB & 95 & 0.050\\
33 & Jamaal Williams & DET & 118 & 0.044\\
34 & Tyler Allgeier & ATL & 91 & 0.035\\
\hline
\end{tabular}
\end{table}

We note that these are just examples of player evaluation metrics provided by
our framework based on a nine-week sample of tracking data,
and there are certainly more quantities to be explored.
Since our framework enables hypothetical comparison at different time
points throughout a play, one could consider other summary functions
such as the sum or average across different time windows (e.g., after
first contact, before crossing the line of scrimmage, etc.)

\section{Discussion}
\label{sec:discussion}

In this paper, we introduce a generative modeling approach for
evaluating the frame-level player movement
via hypothetical comparison. Specifically, we explicitly model the
total movement as a function of a player's step length and turn angle at
every moment within a play. Our models incorporate spatial features
derived from high-resolution tracking data along with player random
effects, which gives us insights into the step-and-turn movement
profiles of NFL players. Since the multilevel models are fit in a Bayesian
framework, we show how they can be used to perform simulation to
generate hypothetical steps at any given frame within a play. This, in
turn, facilitates player evaluation through hypothetical comparison,
which we demonstrate with the expected ending yard line by
ball carriers. While our analysis focuses on nine weeks of data from the
2022 NFL season, our general framework can be applied to a longer time
window or multiple seasons, provided that a larger sample of tracking
data is available.

We highlight that our methods naturally extend to other football
contexts and sports involving player tracking data. For example, in
soccer, player trajectories are driven not only by a single movement
objective
but also by
discrete actions by the player in possession of the ball such as
passing, dribbling, and shooting \citep{fernandez2021framework}. This
can be formalized by appending our step selection framework to an action
selection layer. In particular, at each frame \(t\), a player first
selects an action \(a_t\) among all potential actions \(\mathcal A\)
according to an action model \(p(a_t \mid \boldsymbol {X_t})\). The
resulting movement, conditional on the chosen action \(a_t\), is then
model via a step selection distribution
\(p(s_{t+1}, \varphi_{t+1} \mid a_t, \boldsymbol {X_t})\), which governs
the step length and turn angle for the next frame. This approach enables
hypothetical evaluation of both action choices and player movement.
Specifically, it is possible to evaluate whether a player's observed
action is advantageous relative to alternative available actions in the
same situation, or quantify how different movement paths following the
same action might have altered downstream outcomes.

Further, there is more work to be done regarding our simulation strategy.
Our current framework employs a step selection
simulation strategy to obtain a distribution of hypothetical next steps
at each time point. Moving beyond this setting, we could use our
proposed step and turn models to perform forward trajectory simulation
to generate full ball carrier paths. Specifically, starting from an
initial handoff location, we could repeatedly draw posterior predictive
step length and turn angle, and update spatial covariates needed to
simulate the next frame. This would produce a distribution of entire
hypothetical trajectories rather than local alternative steps at each
moment within a play. Two key challenges arise in pursuing this
extension. First, forward simulation requires a well-calibrated tackle
probability model to govern the termination of simulated trajectories.
Second, it would be necessary to model not only the ball carrier's
movement but also the movements of other players on offense and defense.
Because the on-field spatial configuration at each frame is jointly
determined by the locations and trajectories of all players, holding
non-ball carrier trajectories fixed would introduce systematic bias in
the simulation results. We look forward to exploring these extensions in
order to establish more innovative approaches for player evaluation in sports with tracking data.

\section*{Acknowledgments}
We thank the organizers of the NFL Big Data Bowl 2025 for hosting the competition and providing access to the data.

\section*{Code availability}
All code related to this paper is available at \url{https://github.com/qntkhvn/nfl-step-turn}. The data provided by the NFL Big Data Bowl 2025 are available at \url{https://www.kaggle.com/competitions/nfl-big-data-bowl-2025/data}.

\bibliographystyle{apalike}
\bibliography{references}

\appendix

\section{Assesments of step length model}\label{sec:step-assess}

To model step length, we first attempt to use a distribution suited for
modeling a positive, continuous response. Several candidates here
include Gamma, log-normal and Weibull, as these are also common choices
for modeling step length in the animal movement literature
\citep{hooten2017animal}. However, we find that these models do not
provide a good fit to the data (see our assessment below).
We then consider a transformation for the step
length values. Upon investigation, we choose a scaled arcsine
transformation. In particular, we first normalize the step length to be
between 0 and 1, before applying an arcsine transformation to map the
normalized values to the real line.

\begin{figure}[H]

{\centering \includegraphics[width=0.6\linewidth]{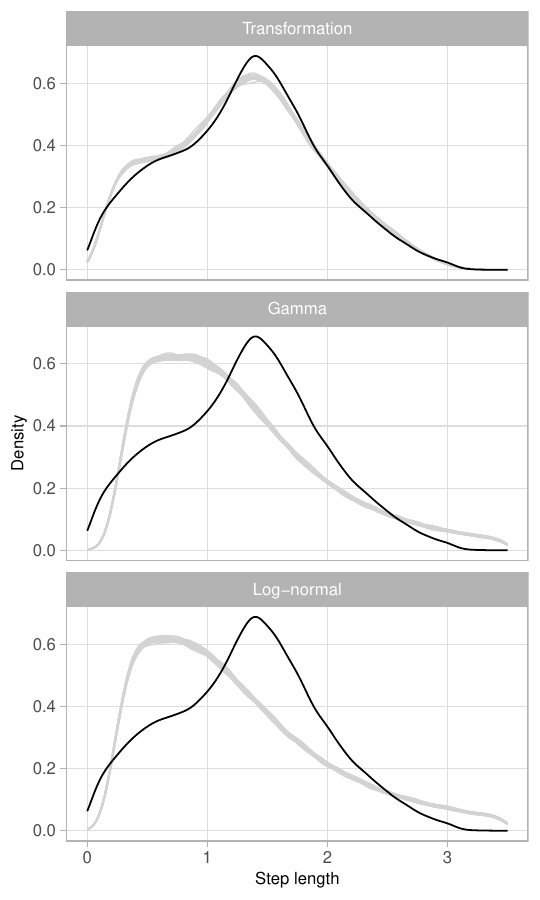} 

}

\caption{Posterior predictive distributions (in gray) and observed distribution (in black) for transformation, Gamma, and log-normal models for step length.}\label{fig:ppc}
\end{figure}

We now perform posterior predictive checks to compare
our proposed step length model with a Gamma model and a log-normal model.
In all models, we control for the same features
and random effects as described in Section \ref{sec:step-model}. Each
model is fit in a Bayesian framework with the same specifications as
before: 4 parallel chains, each made up of 5,000 iterations, and a
warmup of 2,500 draws. After fitting, we generate 25 replications of the
step length values from the posterior predictive distribution and
compare them to the actual data. Figure \ref{fig:ppc} shows density
curves of the posterior predictive distributions for our considered
models along with the observed step length distribution.

We see that the Gamma and log-normal models both provide a poor fit to
the data. In particular, there is a clear departure from the observed
distribution, as we observe a major overestimation of density near
shorter steps and underestimation in the tail region. As a result, the
posterior predictive distributions for these models do match the
empirical step length distribution.
On the other hand, our proposed step length model (via a scaled
arcsine transformation) appears to align most closely with the observed
step length distribution, providing the best overall fit among the three
model candidates. Despite a minor deviation toward the left side (where
the density of shorter step length values are slightly overestimated),
the model generally aligns well with the data across most of the range,
particularly in the right tail. This suggests that our step length model
in (\ref{eq:step}) adequately reflects the key features of the observed
distribution.

Note that in our analysis, the player-level random effect estimates for \(u_j\) are on a transformed scale, resulting from the aforementioned scaled arcsine transformation.
Because this transformation is monotonically increasing, 
higher estimates of \(u_j\) directly correspond to longer step lengths, and vice versa.

\newpage

\section{MCMC diagnostics}
\label{mcmc-convergence}

\begin{figure}[H]

{\centering \includegraphics[width=0.75\linewidth]{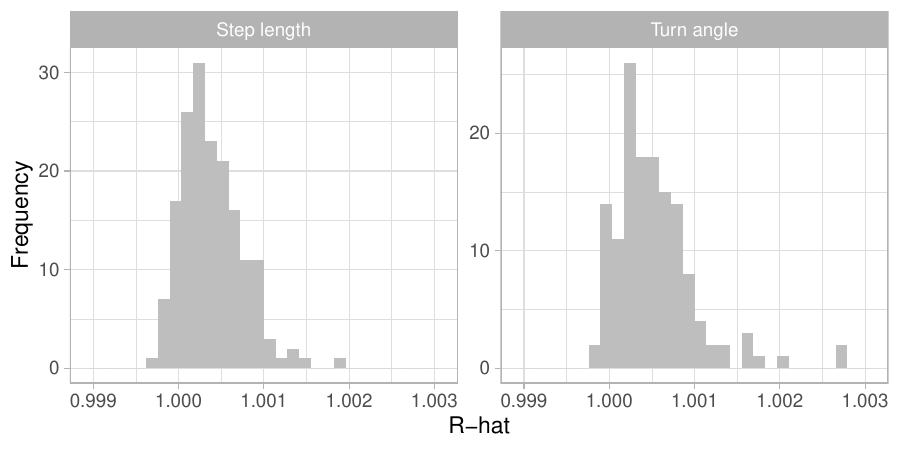} 

}

\caption{Histograms of $\hat{R}$ values for step length and turn angle model parameters. All values are close to 1.}\label{fig:mcmc_rhat}
\end{figure}

\begin{figure}[H]

{\centering \includegraphics[width=0.75\linewidth]{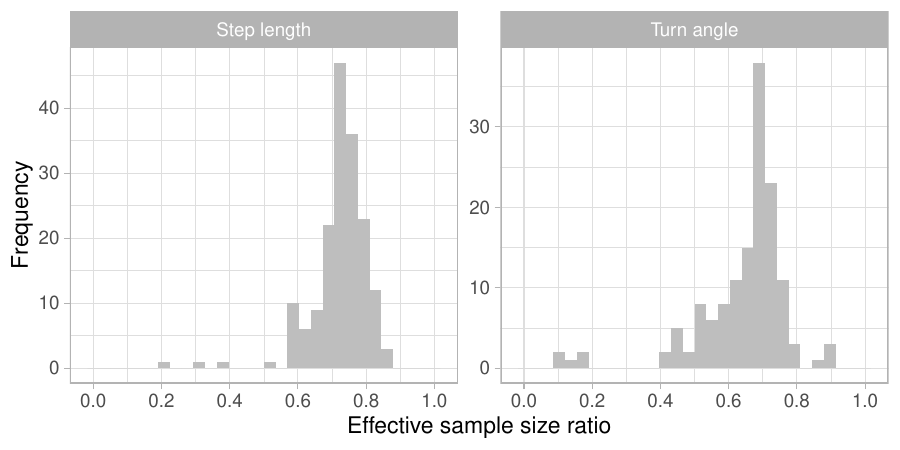} 

}

\caption{Histograms of effective sample size ratio values for step length and turn angle model parameters.}\label{fig:mcmc_ess}
\end{figure}

For both step length and turn angle models, we assess MCMC convergence using \(\hat{R}\) statistics and visual inspection of trace plots.
All parameters show satisfactory \(\hat{R}\) values of near 1 (see Figure \ref{fig:mcmc_rhat}), and their trace plots reveal good mixing and no apparent lack of convergence.
In addition, Figure \ref{fig:mcmc_ess} displays effective sample sample size ratio for all model parameters, showing no issues with sampling efficiency.

\begin{figure}[H]

{\centering \includegraphics[width=0.9\linewidth]{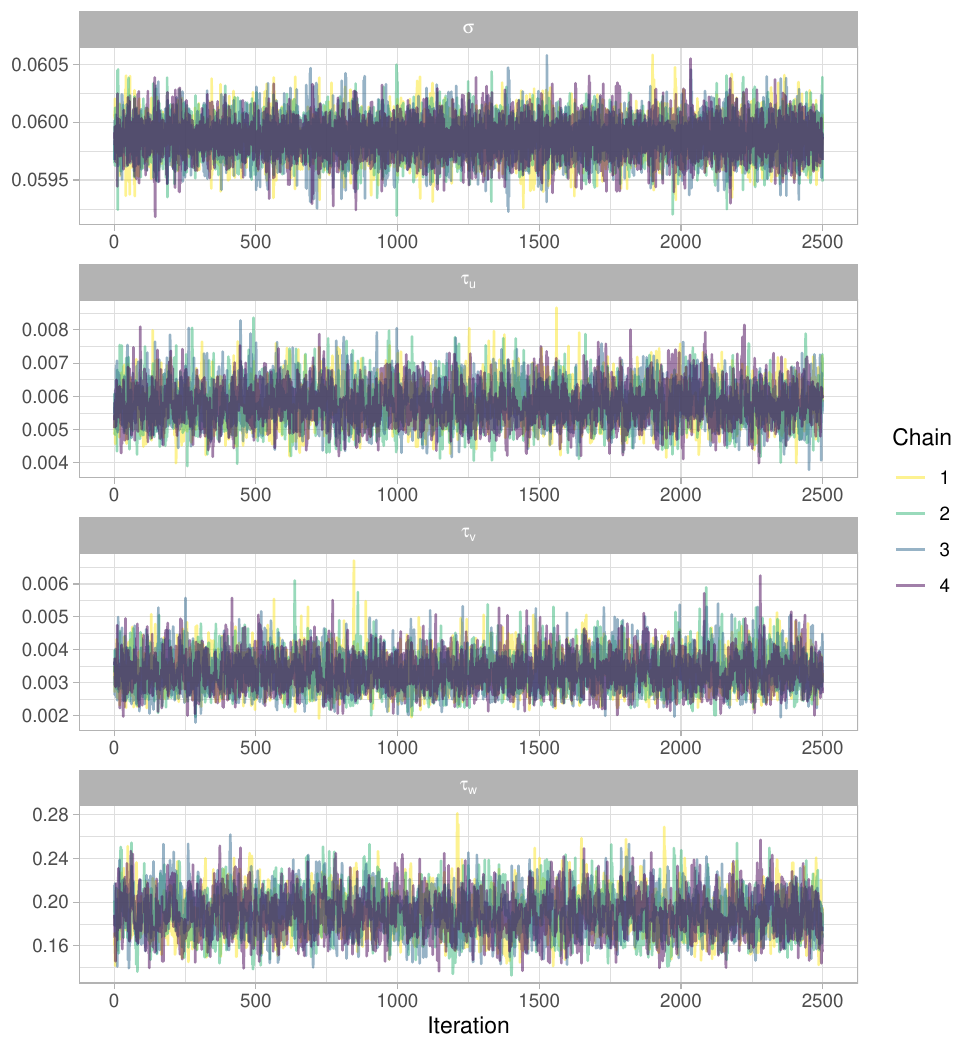} 

}

\caption{Trace plots for the standard deviation parameters in the step length and turn angle models (\(\sigma, \tau_u\), \(\tau_v\), and \(\tau_w\)).}\label{fig:mcmc_trace}
\end{figure}

Below, we present trace plots for the standard deviation parameters (\(\sigma, \tau_u\), \(\tau_v\), and \(\tau_w\)) in Figure \ref{fig:mcmc_trace}.
We focus on these parameters because variance components characterize between-player heterogeneity and are often challenging to estimate in hierarchical models.
Trace plots for the remaining parameters, which also indicate good convergence, are omitted for brevity.

\newpage

\section{Sensitivity analysis}\label{sensitivity}

We investigate the sensitivity of our analysis with respect to the choice of prior distributions. Specifically, we explore different weakly informative prior distributions for the variance components of our step length and turn angle models, namely the standard deviation parameters \(\sigma\), \(\tau_u\), \(\tau_v\), and \(\tau_w\). 
We compare half-Cauchy(0,1), half-normal(0,1), and half-\(t_3\) priors.
For each prior specification considered, we assign the same prior distribution to all standard deviation parameters when fitting the models.

\begin{figure}[H]

{\centering \includegraphics[width=0.9\linewidth]{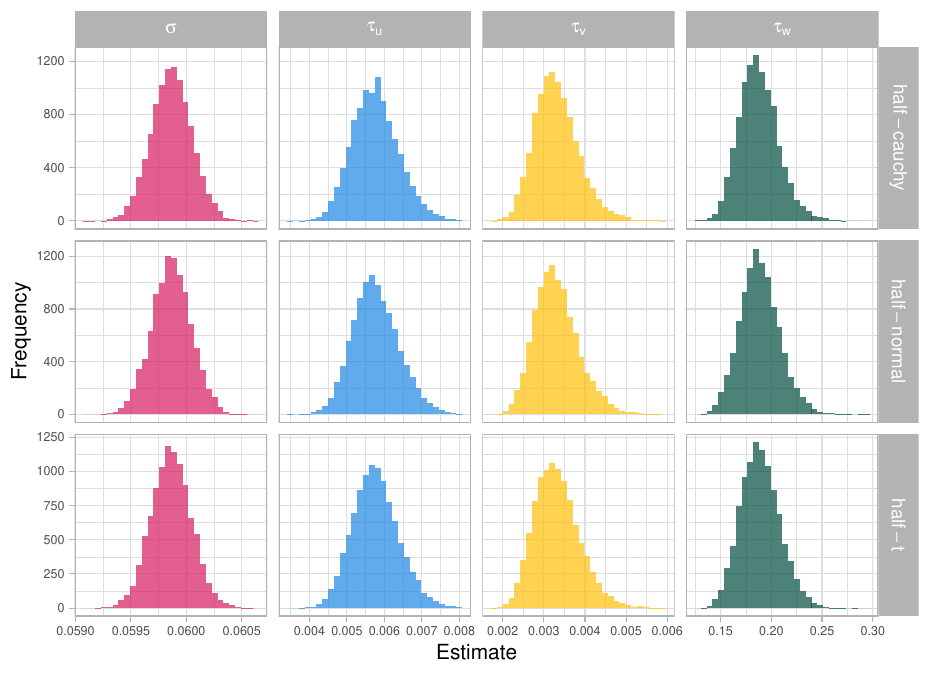}

}

\caption{Comparison of standard deviation parameter estimates for step length and turn angle models with different priors (half-$t_3$, half-normal$(0, 1)$, half-Cauchy$(0, 1)$) for the
standard deviation parameters.}\label{fig:sd_sensitivity}
\end{figure}

Figure \ref{fig:sd_sensitivity} shows the results for our sensitivity analysis.
Overall, the posterior distributions of the standard deviation parameters are highly similar,
and our findings are robust across different prior choices.
After more inspection, we find that the resulting player random effect estimates and rankings are unchanged regardless of the specification.
This suggests that the inferred differences in player movement characteristics are primarily driven by the observed tracking data rather than by the choice of prior.

\newpage

\section{Player ratings: step length
model}\label{step-ratings}

\begin{figure}[H]

{\centering \includegraphics[width=0.6\linewidth]{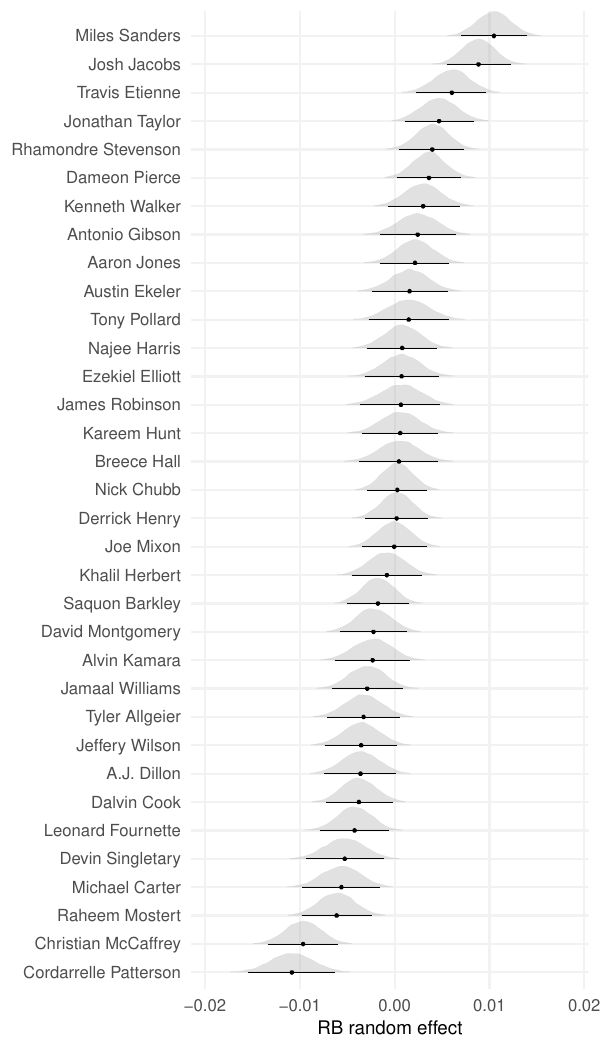} 

}

\newpage

\caption{Posterior distributions of the ball carrier random effect $u_j$ in the step length model (on a transformed scale) for NFL running backs with at least 70 rush attempts on running plays over the first nine weeks of the 2022 regular season. For each player, the posterior mean and corresponding 95\% credible interval are depicted.}\label{fig:step_posterior_distributions}
\end{figure}

\newpage

\section{End-of-play yard line model}\label{yards-model}

We train a multinomial classification model using the CatBoost library to estimate the expected ending yard line $\ell_{ijt}$ for player $j$ at frame $t$ of play $i$.
Our model accounts for frame-level features about the ball carrier and all other players on both offense and defense as summarized in Table
\ref{tab:features}.
It is trained on all data from the provided tracking sample, covering the first nine weeks of the 2022 NFL season.

We perform hyperparameter tuning using grouped 5-fold cross-validation, where observations are partitioned into folds at the game level.
The hyperparameter search space includes \{100, 200, 500, 1000, 1500\} iterations and learning rates of \{0.01, 0.03, 0.05\}, while the maximum tree depth is fixed at 6 in advance.
We find that the combination of 1000 iterations, a learning rate of
0.03, and a maximum tree depth of 6 achieves the best performance in terms of RMSE.

\section{Alternative simulation procedure}\label{alt-sim}

We consider an alternative simulation setting and evaluate a ball
carrier's movement against a distribution of 
hypothetical steps generated from their player-specific random effect.
To do so, we slightly modify our simulation
procedure presented in Section \ref{sec:simulating}. Specifically, to simulate
step length, instead of drawing a new random effect
\(u_{\textsf{new}}^{(h)} \sim \mathcal{N}(0, \tau_u^{2^{(h)}})\) to
represent an average player, we condition on the specific player's
random effect \(u_j^{(h)}\). Similarly, to simulate turn angle, we
replace the generic turn angle random effect draw
\(w_{\text{new}}^{(h)}\) with \(w_j^{(h)}\). Since the player's
estimated random intercept is used, the simulated steps and turns
reflect hypothetical movements based on their own profile rather than a
generic baseline.

\begin{figure}[H]

{\centering \includegraphics[width=0.8\linewidth]{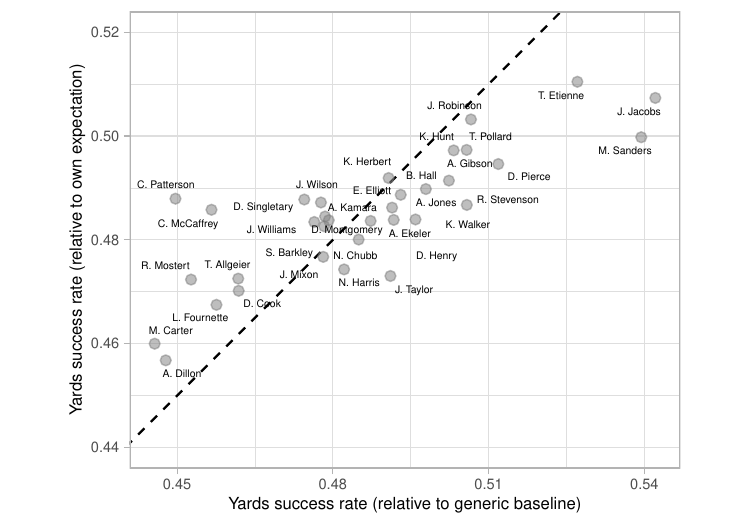} 

}

\caption{Relationship between yards success rate relative to a generic baseline and yards success rate relative to a player's own movement profile.}\label{fig:ysr_comparison}
\end{figure}

We now compute the yards success rate and explosiveness measures (as
described in Examples 1 and 2) to compare our paper's simulation
approach and the alternative strategy. Figure \ref{fig:ysr_comparison}
shows a strong positive correlation,
suggesting that the
yards success rate relative to a player's own movement profile and an
average baseline are well-aligned with each other. However, we observe
observations both above and below the identity line. This is consistent
with the set up of our simulation approaches. In particular, some ball
carriers exhibit higher success rates against generic alternatives than
against their own simulated distributions. Our results also highlight
unique running styles for players like Christian McCaffrey, whose
movement as a patient runner is highly effective relative to his own
expectation, but is penalized when compared to a generic baseline.

\begin{figure}[H]

{\centering \includegraphics[width=1.05\linewidth]{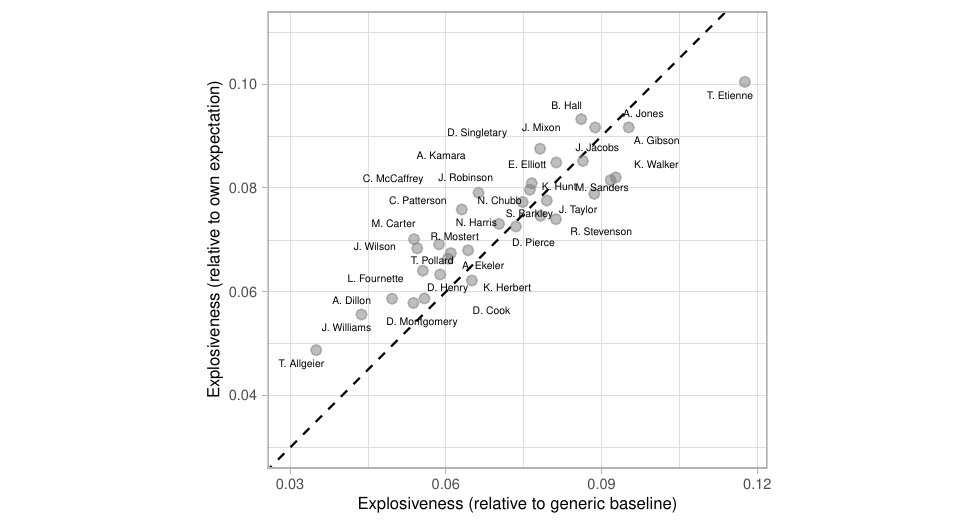} 

}

\caption{Relationship between explosiveness relative to a generic baseline and explosiveness relative to a player's own movement profile.}\label{fig:explosiveness_comparison}
\end{figure}

As for explosiveness, Figure \ref{fig:explosiveness_comparison} displays
a strong relationship
when computing this measure for the
two simulation procedures. Although the two settings generally produce similar evaluations, some ball carriers appear more explosive relative to the generic baseline than to individual-specific distributions (e.g., Travis Etienne).
This highlights how the choice of simulation approach can affect the assessment of player movement profiles.

\end{document}